\documentclass[a4paper,10pt]{article}
\usepackage[utf8x]{inputenc}
\usepackage{graphicx}
\usepackage{comment}
\usepackage{geometry}
\usepackage{url}
\usepackage{amsfonts}
\usepackage{hyperref}

\begin{document}

\begin{center}
\noindent
{\LARGE Keplerian motion of particles with permanent electric

\vskip 0.1cm
\noindent
dipole within cylindrical coaxial electrodes}

\bigskip

\bigskip

Michal \v{S}pa\v{c}ek and Vojt\v{e}ch Petr\'{a}\v{c}ek

\textit{Department of Physics, Faculty of Nuclear Sciences and Physical Engineering, Czech Technical University in Prague, B\v{r}ehov\'{a} 7, 115 19 Prague 1, Czech Republic}

\bigskip

19.2.2019
\end{center}

\bigskip
\bigskip
\bigskip

A research on a possibility of trapping a particle with permanent electric dipole in an electrostatic field has been conducted. For cylindrical coaxial electrodes, Keplerian orbits for some particles were revealed. The exact criterion of successful trapping on a closed orbit within the electrodes is expressed in dimensionless parameters. For more complicated cases where the exact solution is unknown, a useful tool for numerical solution -- local field map interpolation with continuous first derivatives -- is constructed.

\section{Permanent electric dipole -- simplified dynamics}

It has been suggested \cite{1, 2} that for moderate external fields and non-relativistic velocities, the dynamics of a particle with a permanent electric dipole\footnote{A hydrogen atom in o proper quantum state, a water or ozone molecule etc.} $ \vec \mu_e $ can take rather simple form. If the dipole of a particle at $ \vec R $ is considered to align with the local electric intensity $ \vec E = \vec E \left( \vec R \right) $ promptly, the torque equation is effectively eliminated, and what is left is a force on a point-like body. As a matter of fact, it is not the whole dipole what aligns with $ \vec E $ but its projection

$$ \vec \mu_{e(\vec E)} = \mu_{e(\vec E)} \frac{\vec E}{E} = \left( \vec \mu_e \cdot \vec E \right) \cdot \frac{\vec E}{E^2}, \eqno(1.1) $$

\noindent
where $ E \equiv | \vec E | $ and $ \mu_{e(\vec E)} $ denotes the projection's magnitude which is quantized. In the case of a hydrogen atom, for instance, these discrete values are

$$ \mu_{e(\vec E)} = \frac{3}{2} e a_0 n p, \eqno(1.2) $$

\noindent
where $ e $ is an elementary charge, $ a_0 \left( = \frac{\hbar c}{\alpha m_e c^2} \right) $ stands for the Bohr's radius and $ n $ and $ p $ are the principal and the so called parabolic quantum numbers respectively \cite{3, 4}\footnote{Moreover, as stated in \cite{4}, the semiclassical interpretation of the alignment is the precession of $ \vec \mu_{e} $ about $ \vec E $.}. The potential energy of the particle is given as

$$ W_p = \mu_{e(\vec E)} E~~~~~\left( = \mu_{e(\vec E)} \left| \vec E \left( \vec R \right) \right| \right), \eqno(1.3) $$

\noindent
where $ W $ stands for energy to distinguish it from the electric field. The force on the particle therefore is

$$ \vec F = - \vec \nabla W_p \eqno(1.4) $$

\noindent
which is apparently (contrary to a simple charged particle) proportional to the field's derivatives $ \frac{\partial E_i}{\partial X_j} $.

\section{Cylindrical electrodes}

Let us have a vacuum region between two coaxial electrodes. In cylindrical coordinates (with $ z $ coordinate along their joint axis), $ r = r_{\mathrm{in}} $ and $ r = r_{\mathrm{out}} $ simply represent the surfaces of the electrodes. The boundary conditions of the region are

$$ \Phi_{\mathrm{in}} = \Phi \left( r_{\mathrm{in}} \right) \hspace{0.20cm} \eqno(2.1\mathrm{a}) $$
$$ \Phi_{\mathrm{out}} = \Phi \left( r_{\mathrm{out}} \right). \eqno(2.1\mathrm{b}) $$

\noindent
The highly symmetrical setup -- $ \frac{\partial \Phi}{\partial z} = 0 $ and $ \frac{\partial \Phi}{\partial \varphi} = 0 $ -- reduces the Poisson equation $ \Delta \Phi = 0 $ to

$$ 0 = \frac{\partial^2 \Phi}{\partial r^2} + \frac{1}{r} \cdot \frac{\partial \Phi}{\partial r}, \eqno(2.2) $$

\noindent
the unique solution to which (with the boundary conditions (2.1a) and (2.1b) applied) is

$$ \Phi = \frac{\Phi_{\mathrm{out}} \cdot \ln \frac{r}{r_{\mathrm{in}}} - \Phi_{\mathrm{in}} \cdot \ln \frac{r}{r_{\mathrm{out}}}}{\ln \frac{r_{\mathrm{out}}}{r_{\mathrm{in}}}}. \eqno(2.3) $$

\noindent
Since it only depends on $ r = \sqrt{X^2 + Y^2} $, the electric field in cylindrical coordinates follows as

$$ \vec E = \left( E_r, E_{\varphi}, E_z \right) = \left( - \frac{\partial \Phi}{\partial r}, - \frac{1}{r} \cdot \frac{\partial \Phi}{\partial \varphi}, - \frac{\partial \Phi}{\partial z} \right) = \left( - \frac{\Phi_{\mathrm{out}} - \Phi_{\mathrm{in}}}{\ln \frac{r_{\mathrm{out}}}{r_{\mathrm{in}}}} \cdot \frac{1}{r}, 0, 0 \right) \eqno(2.4) $$

\noindent
which implies

$$ E = \left| \vec E \right| = \left| \frac{\Phi_{\mathrm{out}} - \Phi_{\mathrm{in}}}{\ln \frac{r_{\mathrm{out}}}{r_{\mathrm{in}}}} \right| \cdot \frac{1}{r} \eqno(2.5) $$

\noindent
and eventually

$$ W_p = \mu_{e(\vec E)} E = \mu_{e(\vec E)} \left| \frac{\Phi_{\mathrm{out}} - \Phi_{\mathrm{in}}}{\ln \frac{r_{\mathrm{out}}}{r_{\mathrm{in}}}} \right| \cdot \frac{1}{r}. \eqno(2.6) $$

\noindent

Since the projections of $ \mu_{e(\vec E)} $ can be both positive and negative, there also can always be particles with potential energy

$$ W_p = - \frac{\tilde \kappa}{r}, \tilde \kappa > 0 \eqno(2.7) $$

\noindent
which formally matches the potential energy leading to Keplerian orbits.

\section{Kepler-like motion}

Potential energy (2.7) depends on (cylindrical) radial distance $ r $ from the $ z $ axis. Ordinary Kepler potential has a point-like source whereas the force $ \vec F = - \vec \nabla W_p = \left( - \frac{\tilde \kappa}{r^2}, 0, 0 \right) $ always points perpendicular to the $ z $ axis -- the source of the force is a line. Equivalently -- in true Kepler potential, the force $ \vec F $ on a particle, its initial position $ \vec R_0 $\footnote{The origin coincides with the potential centre of symmetry, as usual.} and initial velocity $ \vec v_0 $ lie in one plane each time, but this is not met in the cylindrical region. However, it could be quickly fixed if the initial velocity is decomposed to directions perpendicular and parallel to the $ z $ axis:

$$ \vec v_0 = \vec v_{0(xy)} + \vec v_{0(z)} \eqno(3.1) $$

There is no force in the direction parallel to $ \vec v_{0(z)} $ and the motion is uniform with the velocity $ v_{0(z)} $ that way -- therefore in the frame where the parallel motion vanishes, $ \tilde {\vec v}_0 = \vec v_{0(xy)} $ is in the same plane (perpendicular to $ z $) as $ \tilde{\vec R}_0 = \vec r_0 $ and $ \vec F = \vec F (r) $. The motion is hereby effectively separated into a Keplerian motion in the plane perpendicular to $ z $ and uniform translation of that plane along $ z $.

\begin{figure}[ht]
\begin{center}
\includegraphics[width=10cm]{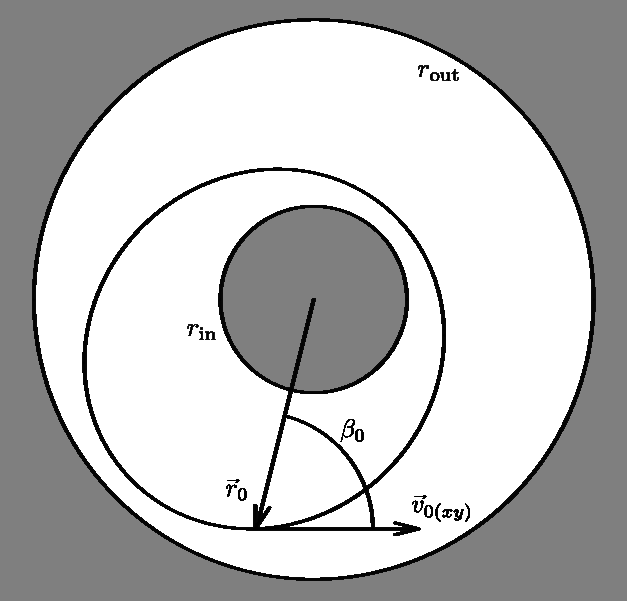}
\caption{Cross-section of the region between coaxial electrodes with initial kinematics of a particle. The plane is perpendicular to $ z $ axis (which is the point in the centre) and moves with the velocity $ v_{0(z)} $ along the axis; $ \vec r_0 $ stands for the particle's initial position, $ \vec v_{0(xy)} $ is its initial velocity and $ \beta_0 $ denotes the angle between the two vectors ($ \beta_0 = 0 $ implies motion towards the $ z $ axis, $ \beta = \pi $ away from). The bodies of the electrodes beyond their surfaces ($ r = r_{\mathrm{in}} $ and $ r_{\mathrm{out}} $) are in grey colour. An example of the closed elliptic orbit of a successfully trapped particle is sketched.}
\label{Ellipse3g}
\end{center}
\end{figure}

For $ v_{0(z)} = 0 $, the angular momentum $ L = L_0 $ is conserved; it can be expressed with radial distance, velocity and direction in the initial time:

$$ L = M v r \sin \beta = M v_{0(xy)} r_0 \sin \beta_0 = L_0; \eqno(3.2) $$

\noindent
$ M $ stands for the particle's mass and $ \beta_0 \in \left< 0, \pi \right> $ is the angle between initial position vector $ \vec r_0 $ and initial velocity $ \vec v_{0(xy)} $. Representation of the quantities is provided in the Figure 1.

Adapted in this way, the problem is straightforward as in any textbook on mechanics and results in an equation of the trajectory in cylindrical coordinates

$$ \frac{\frac{L_0^2}{m \tilde \kappa}}{r} = 1 + \sqrt{1 + \frac{2 W_0 L_0^2}{m \tilde \kappa^2}} \cdot \cos \varphi, \eqno(3.3) $$

\noindent
where $ W_0 = W_{k0} + W_{p0} = W_{k0} - \left| W_{p0} \right| $ denotes the initial total energy\footnote{The form with absolute values holds for $ W_p < 0 $ which only leads to closed orbits.} ($ W_{k0} = \frac{1}{2} M v_{0(xy)}^2 $) and the square root is the conic sections' eccentricity\footnote{Only $ 0 \le e < 1 $ can result in particle's successful trapping.}.

\section{Criterion of trapping}

There are many parameters of the motion -- mass $ M $, initial position $ r_0 $, initial velocity $ v_{0(xy)} $, initial direction of motion $ \beta_0 $, electrodes' parameters $ \Phi_{\mathrm{in}} $, $ r_{\mathrm{in}} $, $ \Phi_{\mathrm{out}} $, $ r_{\mathrm{out}} $, dipole projection $  $ (in case of a hydrogen atom given by two quantum numbers) -- at least nine of them. They can be, however, effectively reduced to four dimensionless quantities to test whether the particle stays inside the region for an unlimited time or not: these are the already introduced $ \sin \beta_0 $ and

$$ \rho \equiv \frac{r_{\mathrm{in}}}{r_0} \left( < 1 \right), \hspace{0.25cm} \eqno(4.1\mathrm{a}) $$
$$ R \equiv \frac{r_{\mathrm{out}}}{r_0} \left( > 1 \right), \hspace{0.15cm} \eqno(4.1\mathrm{b}) $$
$$ \eta_0 \equiv \frac{| W_{p0} |}{W_{k0}} \left( > 0 \right). \eqno(4.1\mathrm{c}) $$

\noindent
Only particles with $ W_p < 0 $ are taken into account. Equation (3.3) transforms into

$$ \frac{\frac{2 r_0}{\eta_0} \sin^2 \beta_0}{r} = 1 + \sqrt{1 + 4 \left( \frac{1}{\eta_0^2} - \frac{1}{\eta_0} \right) \sin^2 \beta_0} \cdot \cos \varphi. \eqno(4.2) $$

\noindent
The fundamental condition for successful trapping is $ r \in \left( r_{\mathrm{in}}, r_{\mathrm{out}} \right) $ or $ r_{\mathrm{in}} < r_{\mathrm{min}} \le r_{\mathrm{max}} < r_{\mathrm{out}} $ where $ r_{\mathrm{min}} $ and $ r_{\mathrm{max}} $ are the ellipse's\footnote{Circle, as a special case, is also included.} apsides. In terms of $ R $, $ \rho $, $ \eta_0 $, $ \sin \beta_0 $ follows

$$ \rho < \frac{\frac{2}{\eta_0} \sin^2 \beta_0}{1 + \sqrt{1 + 4 \left( \frac{1}{\eta_0^2} - \frac{1}{\eta_0} \right) \sin^2 \beta_0}}, \eqno(4.3\mathrm{a}) $$
$$ R > \frac{\frac{2}{\eta_0} \sin^2 \beta_0}{1 - \sqrt{1 + 4 \left( \frac{1}{\eta_0^2} - \frac{1}{\eta_0} \right) \sin^2 \beta_0}}. \eqno(4.3\mathrm{b}) $$

If the square root in the first relation is expressed and then both sides raised to the second power, a simple inequality is obtained:

$$ \rho^2 - \rho^2 \eta_0 < \sin^2 \beta_0 - \eta_0 \rho. \eqno(4.4) $$

\noindent
This could be understood in two ways:

$$ \rho^2 - \rho^2 \eta_0 + \eta_0 \rho < \sin^2 \beta_0 \le 1 \eqno(4.5\mathrm{a}) $$
$$ \eta_0 < \frac{1}{\rho ( 1 - \rho )} \sin^2 \beta_0 - \frac{\rho}{1 - \rho}. \eqno(4.5\mathrm{b}) $$

\noindent
The first implies $ \eta_0 \le 1 + \frac{1}{\rho} $, while the second one puts a restriction on initial conditions in terms both $ \eta_0 $ and $ \sin \beta_0 $:

$$ \eta_0 < \frac{1}{\rho ( 1 - \rho )} \sin^2 \beta_0 - \frac{\rho}{1 - \rho} \eqno(4.6) $$

\noindent
The same procedure applied on (4.3b) results in $ \eta_0 \ge 1 + \frac{1}{R} $ and

$$ \eta_0 > - \frac{1}{R (R - 1)} \sin^2 \beta_0 + \frac{R}{(R - 1)}. \eqno(4.7) $$

\noindent
The requirement of $ 0 \le e < 1 $ does not bring any more restrictions. The criterion of a particle being trapped on a closed orbit within the electrodes is therefore given by (4.6), (4.7) and naturally $ 0 \le \sin \beta_0 \le 1 $ and represents an area in $ \sin \beta_0 $-$ \eta_0 $ diagram as depicted in the Figure 2.

\begin{figure}[ht]
\begin{center}
\includegraphics[width=12cm]{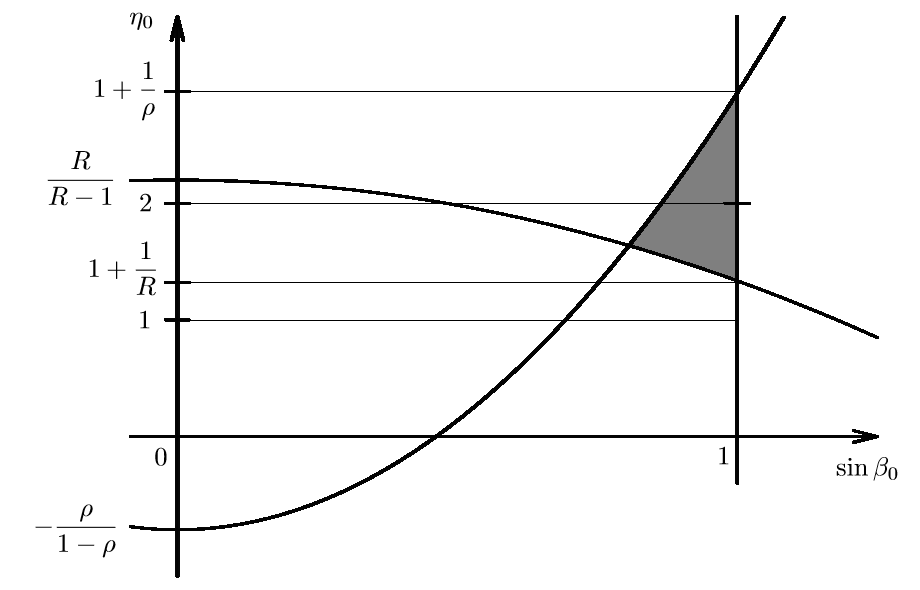}
\caption{Graphical representation of the criterion of particle preservation on a closed trajectory inside the cylindrical electrostatic trap -- successful trapping is the grey area in the $ \sin \beta_0 $--$ \eta_0 $ diagram; $ \eta_0 \equiv \frac{| E_{p0} |}{E_{k0}} $, $ \beta_0 \equiv \angle \left( \vec r_0, \vec v_{0(xy)} \right) $, $ R \equiv \frac{r_{\mathrm{out}}}{r_0} $, $ \rho \equiv \frac{r_{\mathrm{in}}}{r_0} $ (also in Figure 1). The area is defined by three inequalities: $ \sin \beta_0 \le 1 $, $ \eta_0 > - \frac{1}{R (R - 1)} \sin^2 \beta_0 + \frac{R}{(R - 1)} $ (bordered by the decreasing parabola) and $ \eta_0 < \frac{1}{\rho ( 1 - \rho )} \sin^2 \beta_0 - \frac{\rho}{1 - \rho} $ (the increasing parabola as a boundary). The $ ( 1, 2 ) $ point (marked with a cross on the right-hand side of the grey area) represents the circle orbits. For a narrowing space between the electrodes $ r_{\mathrm{in}} \rightarrow r_{\mathrm{out}} $, the three apexes of the grey area converge exactly to the $ (1, 2) $ point and for $ r_{\mathrm{in}} < r_{\mathrm{out}} $ there always is a space left for a circle orbit.}
\label{Trap1e}
\end{center}
\end{figure}

\section{Trap specifics beyond Keplerian potential}

The parameters $ \eta_0 $ and $ \sin \beta_0 $ do not describe the state of a particle in a unique way -- for a trapped particle with given $ r_0 $, $ v_{0(xy)} $ and $ \sin \beta_0 $, there are up to four distinct $ \vec v_{0(xy)} $. It should also be noted that $ \rho $, $ R $ and $ \eta_0 $ are not fully independent -- for example, radial pulsation of the inner electrode surface also changes the potential which is not the case in gravity.

If a particle with a permanent dipole is inserted between the electrodes through the external electrode\footnote{The same holds for the internal, of course.} non-tangentially ($ r_0 = r_{\mathrm{out}} \Rightarrow R = 1 $, $ \sin \beta_0 \neq 1 $), it surely hits one of the electrode elsewhere. The trap is therefore suitable for particles produced inside it -- for example through charge exchange of ions which obey Lorentz force\footnote{The motivation of the analysis is connected to production of antihydrogen or Rydberg matter.} -- or for particles transported along the $ z $ axis. Appropriate perturbation of the electrodes' boundaries $ r = r_{\mathrm{const}} \rightarrow r = r (z) $, in order to trap the particles in the $ z $ direction as well, has not been sufficiently explored yet.

\section{Numerical approach to electric dipole in electrostatic field}

According to symmetry of the given field and the coordinate system, there are up to nine non-zero functions $ \frac{\partial E_i}{\partial X_j} \left( \vec R \right) $ on which the force on the particle is dependent. If the analytic form of the field is not known, it is often expressed in the form of a field map in discrete points in space, which has to be interpolated to determine the field in any point where the particle is located. For simplicity, let us examine one-dimensional map of a scalar ($ E = E(X) $) in equidistant points $ X_n = n \Delta X, n \in \left \{ N_i, ..., N_f \right\}, N_i, N_f \in \mathbb{Z} $. Splining the whole map is both lengthy and error-prone, it should be performed locally. However, as it is usually $ \vec E $ what is mapped, simple linear interpolation

$$ E (X) = \frac{E_{n + 1} - E_n}{\Delta X} \left( X - n \Delta X \right) + E_n, \forall X \in I_n. \eqno(6.1) $$

\noindent
does not keep the first derivative $ \frac{\partial E}{\partial X} $ continuous. Since the force on a dipole particle is proportional to first space derivatives of the (from the field map interpolated) electric field, it leads to non-physical results which resemble apsis precession or perturbing particle's plane of orbit; example is in Figure 3.

\begin{figure}[ht]
\begin{center}
\includegraphics[width=10cm]{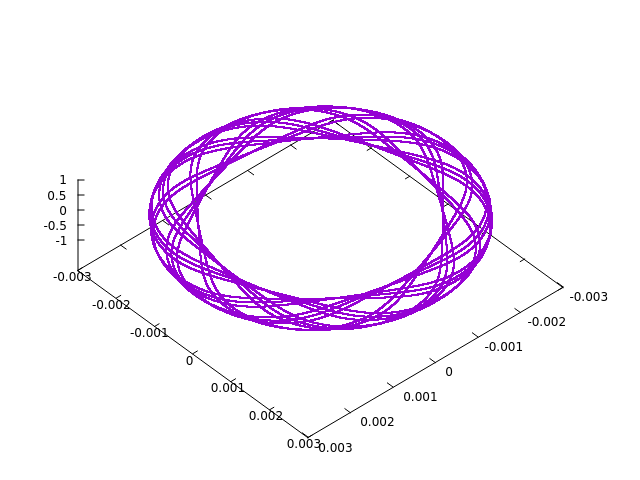}
\caption{Linear interpolation of a field map is too rough for a dipole particle since the force on it in such a field depends on the field's first derivative. The particle follows an elliptic orbit but from a specific time it unexpectedly deviates into motion similar to apsis precession which should not take place -- perspective view.}
\label{Egg}
\end{center}
\end{figure}

A satisfactory solution in $ \left< X_n, X_{n + 1} \right) $\footnote{These are neighbouring points in the map: $ X_{n + 1} - X_n = \Delta X $} is a specific cubic interpolation if not only $ \left( X_n, E_n \right) $ and $ \left( X_{n + 1}, E_{n + 1} \right) $ but also $ \left( X_{n - 1}, E_{n - 1} \right) $ and $ \left( X_{n + 2}, E_{n + 2} \right) $ are applied. In $ \left< X_n, X_n + \frac{\Delta X}{4} \right) $ a straight line\footnote{The lines right in $ \left< X_n, X_n + \frac{\Delta X}{4} \right) $ and in $ \left< X_{n + 1} - \frac{\Delta X}{4}, X_{n + 1} \right) $ make the evaluation of parameters in (6.2) less complicated and, moreover, half of the interpolation completely avoids the higher polynomial.} is constructed to pass through the $ \left( X_n, E_n \right) $ point with the slope given by $ \left( X_{n - 1}, E_{n - 1} \right) $ and $ \left( X_{n + 1}, E_{n + 1} \right) $ as $ \frac{X_{n + 1} - X_{n - 1}}{2 \Delta X} $. A line in $ \left< X_{n + 1} - \frac{\Delta X}{4}, X_{n + 1} \right) $ is found likewise. The middle half of the interval is eventually interpolated with a polynomial of a degree up to three which connects to the two lines so that the whole function

$$ E(X) = \left \{ \begin{array}{l} a_L X + b_L, \hspace{1.75cm} X \in \left< X_n, X_n + \frac{\Delta X}{4} \right) \\ a X^3 + b X^2 + c X + d, X \in \left< X_n + \frac{\Delta X}{4}, X_{n + 1} - \frac{\Delta X}{4} \right) \\ a_R X + b_R, \hspace{1.65cm} X \in \left< X_{n + 1} - \frac{\Delta X}{4}, X_{n + 1} \right) \end{array} \right., \eqno(6.2) $$

\noindent
is continuous in $ \left( X_n, X_{n + 1} \right) $, as well as its first derivative. The coefficients in (6.2) are given by simultaneous equations

\newpage

$$ a_L = \frac{X_{n + 1} - X_{n - 1}}{2 \Delta X} \hspace{3.6cm} \eqno(6.3\mathrm{a}) $$
$$ E_n = a_L X_n + b_L \hspace{4.2cm} \eqno(6.3\mathrm{b}) $$
$$ a_L \left( X_{n - 1} + \frac{5 \Delta X}{4} \right) + b_L = a \left( X_{n - 1} + \frac{5 \Delta X}{4} \right)^3 + b \left( X_{n - 1} + \frac{5 \Delta X}{4} \right)^2 \hspace{2.85cm} $$
$$ + c \left( X_{n - 1} + \frac{5 \Delta X}{4} \right) + d \hspace{2.0cm} \eqno(6.3\mathrm{c}) $$
$$ a_R \left( X_{n - 1} + \frac{7 \Delta X}{4} \right) + b_R = a \left( X_{n - 1} + \frac{7 \Delta X}{4} \right)^3 + b \left( X_{n - 1} + \frac{7 \Delta X}{4} \right)^2 \hspace{2.85cm} $$
$$ + c \left( X_{n - 1} + \frac{7 \Delta X}{4} \right) + d \hspace{2.0cm} \eqno(6.3\mathrm{d}) $$
$$ \frac{E_{n + 1} - E_{n - 1}}{2 \Delta X} = 3 a \left( X_{n - 1} + \frac{5 \Delta X}{4} \right)^2 + 2 b \left( X_{n - 1} + \frac{5 \Delta X}{4} \right) + c \hspace{0.4cm} \eqno(6.3\mathrm{e}) $$
$$ \frac{E_{n + 2} - E_n}{2 \Delta X} = 3 a \left( X_{n - 1} + \frac{7 \Delta X}{4} \right)^2 + 2 b \left( X_{n - 1} + \frac{7 \Delta X}{4} \right) + c \eqno(6.3\mathrm{f}) $$
$$ a_R = \frac{X_{n + 2} - X_n}{2 \Delta X} \hspace{4.1cm} \eqno(6.3\mathrm{g}) $$
$$ E_{n + 1} = a_R X_{n + 1} + b_R, \hspace{4.25cm} \eqno(6.3\mathrm{h}) $$

\noindent
the solution of which is

$$ a_L = \frac{X_{n + 1} - X_{n - 1}}{2 \Delta X} \hspace{9.2cm} \eqno(6.4\mathrm{a}) $$
$$ b_L = E_n - \frac{E_{n + 1} - E_{n - 1}}{2 \Delta} \left( X_{n - 1} + \Delta X \right) \hspace{6.2cm} \eqno(6.4\mathrm{b}) $$
$$ a = \frac{4}{(\Delta X)^3} \left( E_{n + 2} - 3 E_{n + 1} + 3 E_n - E_{n - 1} \right) \hspace{5.3cm} \eqno(6.4\mathrm{c}) $$
$$ b = \frac{1}{2 (\Delta X)^2} \left( 3 E_{n + 2} - 7 E_{n + 1} + 5 E_n - E_{n - 1} \right) + \hspace{5.5cm} $$
$$ + \frac{1}{(\Delta X)^3} \left( 12 X_{n - 1} + 19 \Delta X \right) \left( E_{n - 1} - 3 E_n + 3 E_{n + 1} - E_{n + 2} \right) \hspace{2.3cm} \eqno(6.4\mathrm{d}) $$
$$ c = \frac{E_{n + 2} - E_n}{2 \Delta X} + \frac{1}{4 (\Delta X)^2} \left( 4 X_{n - 1} + 7 \Delta X \right) \left( E_{n - 1} - 5 E_n + 7 E_{n + 1} - 3 E_{n + 2} \right) + \hspace{0.9cm} $$
$$ \hspace{1.1 cm} + \frac{1}{4 (\Delta X)^3} \left( 4 X_{n - 1} + 7 \Delta X \right) \left( 12 X_{n - 1} + 17 \Delta X \right) \left( E_{n + 2} - 3 E_{n + 1} + 3 E_n - E_{n - 1} \right) \eqno(6.4\mathrm{e}) $$
$$ d = E_{n + 1} + \left( X_{n - 1} + 2 \Delta X \right) \frac{E_n - E_{n + 2}}{2 \Delta X} + \hspace{6.8cm} $$
$$ + \frac{1}{32 (\Delta X)^2} \left( 4 X_{n - 1} + 7 \Delta X \right)^2 \left( 3 E_{n + 2} - 7 E_{n + 1} + 5 E_n - E_{n - 1} \right) + \hspace{2.8cm} $$
$$ + \frac{1}{16 (\Delta X)^3} \left( 4 X_{n - 1} + 5 \Delta X \right) \left( 4 X_{n - 1} + 7 \Delta X \right)^2 \left( E_{n - 1} - 3 E_n + 3 E_{n + 1} - E_{n + 2} \right) \eqno(6.4\mathrm{f}) $$
$$ a_R = \frac{X_{n +2} - X_n}{2 \Delta X} \hspace{10.2cm} \eqno(6.4\mathrm{g}) $$
$$ b_R = E_{n + 1} - \frac{E_{n +2} - E_n}{2 \Delta X} \left( X_{n - 1} + 2 \Delta X \right). \hspace{6.4cm} \eqno(6.4\mathrm{h})  $$

\noindent
What makes the interpolated derivative continuous even in the field map vertices $ X_n $ is that the derivative from left and right are both given by the same formula and by the same values. Example of an interpolation in two neighbouring intervals is in Figure 4.

\begin{figure}[ht]
\begin{center}
\includegraphics[width=12cm]{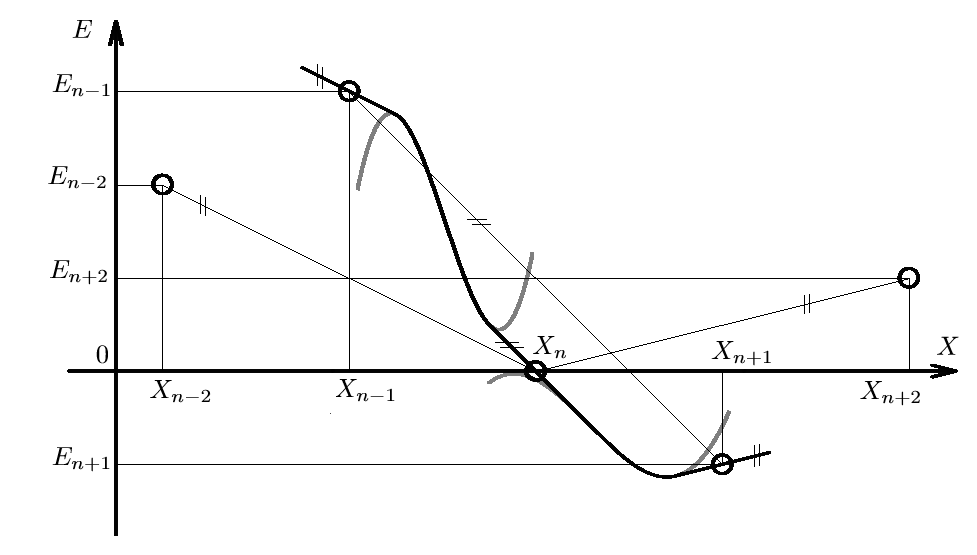}
\caption{Local field map interpolation continuous in each point up to its first derivative. In $ \left< X_{n - 1}, X_n \right) $, the four points $ \left( X_{n - 2}, E_{n - 2} \right) $, $ \left( X_{n - 1}, E_{n - 1} \right) $, $ \left( X_n, E_n \right) $ and $ \left( X_{n + 1}, E_{n + 1} \right) $ are applied -- a line, the slope of which equals the connecting line of $ \left( X_{n - 1}, E_{n - 1} \right) $ and $ \left( X_{n + 1}, E_{n + 1} \right) $ (thin line; parallelism marked), passes through $ \left( X_n, E_n \right) $; similarly for the line passing through $ \left( X_{n - 1}, E_{n - 1} \right) $ with the points $ \left( X_{n - 2}, E_{n - 2} \right) $ and $ \left( X_n, E_n \right) $ applied. The two lines are connected in the middle of the interval $ \left< X_{n - 1}, X_n \right) $ with a polynomial of a degree up to three. The same procedure in $ \left< X_n, X_{n + 1} \right) $ makes the interpolated function with its first derivative continuous even in the $ X_n $ point -- $ \left( X_{n - 1}, E_{n - 1} \right) $ and $ \left( X_{n + 1}, E_{n + 1} \right) $, determining the interpolated derivative, belong to both quartets. The follow-up of the polynomials where they are not considered has a grey colour.}
\label{Interpolation5f}
\end{center}
\end{figure}

For a multivariate function, the interpolation has to be carried out multiple times, first for proper points on lines connecting the map vertices, and subsequently in the given point -- this is the common Particle in Cell algorithm.

\section{Conclusion}

In a model of a point-like torque-less particle with permanent electric dipole, a pair of cylindrical coaxial electrodes was found to act on the particle in a way that it undergoes a Keplerian orbit in a special frame. The exact conditions when the particle stays within the electrodes for an unlimited amount of time has been derived and the result discussed. A local interpolation of a field map, the first derivative of which is continuous, has been constructed and its inevitability for numerical simulations of electric dipole particles in electric field given by a field map was justified.


\begin{thebibliography}{}


\bibitem{1} \v{S}PA\v{C}EK, Michal. \textit{Dynamics of anti-hydrogen motion in the AEGIS experiment}. Prague, 2012. Diploma thesis. Czech Technical University in Prague, Faculty of Nuclear Sciences and Physical Engineering, Department of Physics. Supervised by doc. RNDr. Vojt\v{e}ch Petr\'{a}\v{c}ek, CSc.


\bibitem{2} \v{S}PA\v{C}EK, Michal; PETR\'{A}\v{C}EK, Vojt\v{e}ch. \textit{Internal and external dynamics of antihydrogen inelectric and magnetic fields of arbitrary orientation}. [arxiv.org] Available from: \url{https://arxiv.org/pdf/1206.5171.pdf}.


\bibitem{3} LANDAU, Lev Davidovich; LIFSHITZ, Evgeny Mikhailovich. \textit{Quantum mechanics : Non-relativistic theory}. Third edition, revised and enlarged. [Oxford] : Pergamon Press, 1977. 677 p. ISBN 0-08-020940-8.


\bibitem{4} BORN, Max. \textit{Vorlesungen \"{u}ber Atommechanik}. Berlin: Springer, 1925.


\end{thebibliography}
\end{document}